# Actuality and Future of Optical Systems

Nataša Živić
Institute for Data Communications Systems
University of Siegen
Siegen, Germany
E-mail: natasa.zivic@uni-siegen.de

***Abstract*—Today's high-capacity telecommunication systems are unimaginable without the use of optical fibres and accompanying optical components which are briefly presented in this paper with an accent on the receivers of optical signal. An overview of main characteristics of the key elements and the description of the most popular modulation formats in modern optical systems is shown. Some achievements and the directions of the further development in this area are also mentioned.**

## I. Introduction

The basic elements of an optical system are the source of the optical signal, the optical transmission medium and the receiver. *Light emitting diode* (LED) or laser diode is used as the optical source, whereas PIN or avalanche photodiode is used on the receiving side. The transmission medium can be free space (air) or optical fibre. Two basic types of optical fibres are used: multimode (MM) and single-mode (SM) fibre. Depending on the needed information capacity, distance, price and other factors, the choice of different components is made.

Multimode fibres have much worse characteristics regarding their capacity compared to single-mode ones, but their use is cheaper, as LEDs can be used as sources. That is why MM fibres are used for short distances and when a huge transmission capacity is not necessary (up to 1 Gb/s). MM fibres are most often used in LANs. The systems using free space as transmission medium can operate at distances up to 5 km, on the condition of direct optical visibility between the transmitter and the receiver.

The only transmission medium allowing large digital rates at longer distances is single-mode optical fibre. Only laser diode is used for excitation of SM fibres.

Apart from the basic elements, an optical system can also include optical amplifiers which can be placed along the optical fibre, as well as on transmitting or receiving side (booster, optical pre-amplifier).

## II. Optical Transmitters

A transmitter has a double role – as a light source and as a modulator of the generated light. In digital systems modulation is most often done by changing the light intensity which is launched into the fibre. For the light emission into the single-mode fibre, laser diodes are used at 1310 and 1550 nm, and for the multimode fibre a LED diode at 850 and 1300 nm is used. Earlier, transmitters were made of integrated circuits, discrete elements (resistors, capacitors) and LEDs or laser diodes situated on the same board. Today the optical transmitter is produced as an autonomous hybrid construction comprising integrated circuits and discrete elements. The basic characteristics of the transmitter are the source power, maximal speed of modulation, focus, i.e. diameter of output beam of light, linearity, wavelength, dimensions, price and reliability.

LEDs used in transmitters have simpler structure and generate non-coherent light of less power, and laser diodes are more complex and emit coherent light of larger power. Besides, due to the narrower beam of the emitted light, laser diodes can be more effectively coupled with the optical fibre, but they have much greater speed of modulation due to small spectral width. On the other hand, LEDs are more reliable, have better linearity and lower price, and the current-voltage characteristic is of little dependence on temperature and aging.

"Fig. 1" shows one of the typical realizations of the LED with hetero-structures and active zone emitting light between them. The basic difference between LED and laser diode is that the laser diode has the optical resonator consisting of two surfaces with a large reflection, i.e. two semi-transparent mirrors on the front and rear side of the laser diode.

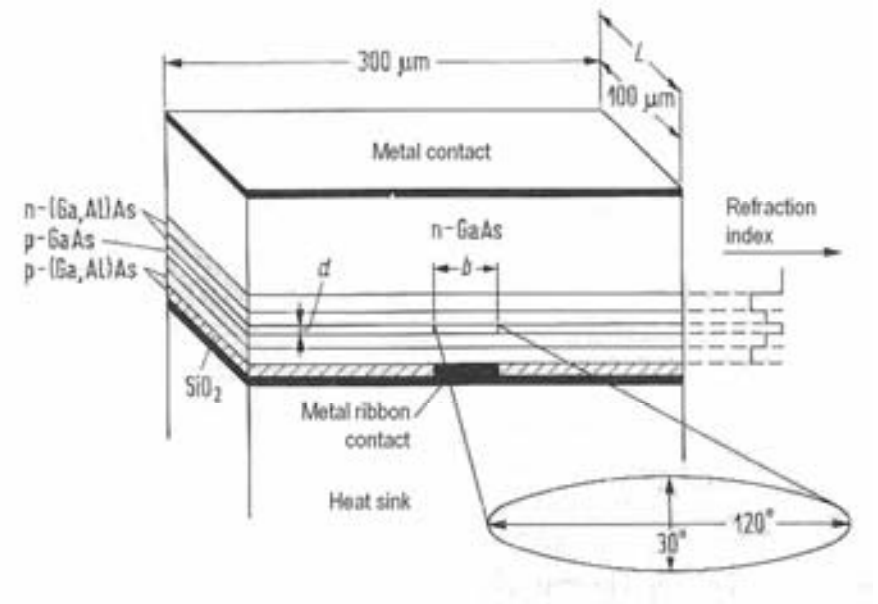

Figure 1. LED with hetero-structures

The first semiconductor laser diodes had simple sandwich structures comprising two layers of GaAs crystals of which one layer is of p-type, the other one of n-type, while the front and the rear side were polished and represent the mirrors of the Perot-Fabry's resonator. Nowadays, we use laser diodes with hetero structures between which there is a narrow active zone (with shortened resonator length from typical 250 to 25μm). Hetero-structures consist of a number of layers with different energy gaps between the valent and the conducting zone and with the different refraction indices. The active zone is made up of semiconductors of p-type with a lower energy gap and a higher refraction index, so that this region acts like a planar waveguide. The thickness of the active zone is below 0.1 μm, and the width is less than 5 μm (narrow-zone lasers) which enables stable work with the transversal type of oscillation.

The amplifier of the electromagnetic wave and the feedback circuit are united in the laser diode. The amplifier is the active zone (with inverse population of the electron energy levels) whereas the feedback is provided by the mirror reflection of the Perot-Fabry's resonator. The basic laser disadvantage with the Perot-Fabry's resonator is multi-mode work regime, and that is why the spectrum of emitted light for this type of laser is much wider than the individual line in the spectrum.

A small spectrum width is needed for digital rates higher than 1 Gb/s, so the lasers were made with the feedback achieved by periodic change of the active zone's refraction indice along the wave propagation axis. With such structures, better selection of the feedback is obtained, and in that way, smaller width of the spectrum line (3 MHz, in relation to the Perot-Fabry's laser where the spectrum width is 100 MHz). There are two basic realizations of such lasers: Distributed Bragg Reflector (DBR), and with a kind of periodic structure, i.e. Distributed Feedback (DFB). Both realizations are shown in "Fig. 2".

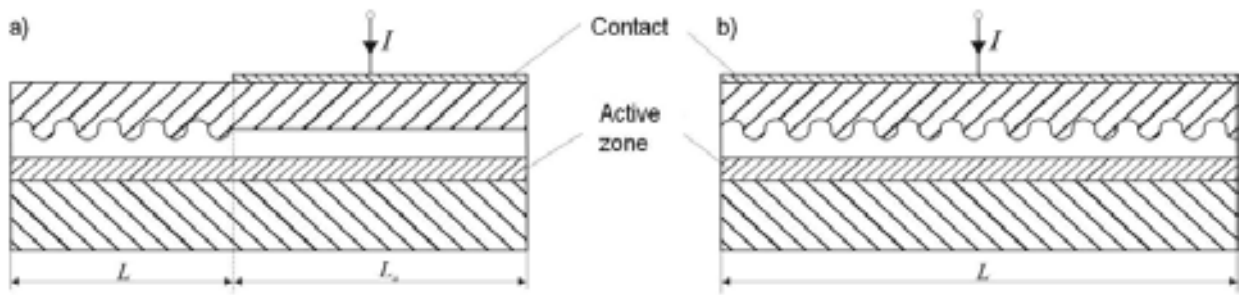

Figure 2. a) DBR laser; b) DFB laser

By DFB laser (which is more developed than DBR), it is possible to get the modulation speed of over 40 Gb/s with much lower threshold current when the laser effect starts (between 10 and 20 mA in room temperature). The disadvantage is that the threshold current increases with the temperature increase and the laser amplification decreases, so it is necessary to achieve good temperature stability in the closest environment of the laser structure. Besides, the temperature change very much affects the wavelength change of the emitted light. This is very important in DWDM (*Dense Wavelength Division Multiplex*) applications, as the distance between the neighboring channels is less than 1 nm.

A huge improvement has been recently made in developing lasers with vertical radiation (Vertical Cavity Surface Emitting Laser, VCSEL). These lasers became attractive due to the low price, low power consumption and easy coupling to the fibre. They are realized on the GaAs substrate above which there is a sandwich structure made up of two Brag gratings (mirrors) and between them there is the active zone. The direction of the emitted light is vertical to the substrate. The main application of these lasers is in LANs and SANs at the wavelength of 850 nm. However, VCSEL for work at 1300 and 1550 nm is also being developed. Now, with these lasers, it is possible to achieve the optical transmission with the 2.5 Gb/s rate over distances longer than 50 km (11), and their further development is to be expected.

## III. Optical Receivers

The function of the optical receiver is converting the signal from optical into the electrical domain, i.e. its demodulation and decision, namely the recognition of the sent series of symbols. The basic component of the receiver in which the conversion of the optical signal into the electric signal is done is the PIN diode. The PIN diode consists of two layers of p and *n*-type, between which there is a (depleted) intrinsic *i*-region. When light comes to this region, having passed through a thin *p*-layer, the generations of pairs of electrons and holes appear, moving in the direction of the electric field, and thus the current is created. The depleted (its own, intrinsic) region is, in fact, slightly doped *p*-type layer which almost does not contain free electrical charge carriers, which lead to a high resistance, and in that way a larger electric field within it. Thus, a more effective creation of the electric charge flow is achieved, than in a common silicon photodiode with *pn*-junction. In working at wavelengths of 1300 and 1550 nm, the most often used are semiconductors made of InGaAsP on InP substrate, or GaAlAsSb on the substrate of GaSb. Germanium photodiodes are also being used (but slightly less effective). For the wavelength range of about 850 nm, silicon photodiodes are used.

The basic characteristics of the optical detector are: conversion coefficient, spectral response, sensibility, response time, dark current, noise (thermal, quantum, dark noise). Conversion coefficient (response) is defined as the relationship of the output current from detector and the exciting optical power, while the spectral response is the curve of the dependence of conversion coefficient versus wavelength. Response time is the time needed for the detector output current to be converted from 10% to 90% of its maximum value by the step change of input optical power. Response time directly affects the bandwidth of the receiver.

The sensitivity of receiver is defined as the minimal optical power (versus digital rate) which can be detected by the receiver, at the given BER value (typical $10^{-9}$). Maximal sensitivity is determined by the quantum limit (QL), i.e. sensitivity when the conversion efficiency of the optical into the electric impulse is 100% and in perfect conditions if there were no thermal and dark noise and with practically infinite bandwidth. The newest PIN diodes reach the quantum limit of -50 dBm, when the BER value is $10^{-9}$ at rate of 10 Gb/s (amplitude modulation is supposed), while their real sensitivity is 20-30 dB above this limit.

With PIN photodiode, each incoming photon generates one electron-hole pair at the most. The PIN diode conversion coefficient is between 0.5 and 0.7 A/W. To increase the number of the absorbed photons, it is necessary for the intrinsic region to be thicker, but thus the response time and the bandwidth of the receiver are reduced, due to the longer passing time of generated carriers of electric charge through this region. With the silicon PIN diode having thickness of 10 μm in the intrinsic region, the bandwidth is 1.59 GHz, which is not at all enough for modern systems with large rates. The maximal bandwidth with today's PIN diodes is over 50 GHz.

The avalanche photodiode (APD) consisting of $p^+ipn^+$ structure with inverse polarization is used to increase the coefficient of the receiver's conversion. Unlike the PIN diode, this structure comprises another layer of p-type between the intrinsic and $n^+$ region which is called the avalanche area. There is a very high electric field in this layer which accelerates the generated electrons made in the intrinsic area. These electrons have kinetic energy high enough to excite other valent electrons which make the avalanche effect. At average, the amplifying factor due to the avalanche effect is from 70-100, which means that one input photon generates 70-100 electron-hole pairs. The amplifying factor (multiplication) is proportional to the inverse polarization voltage.

The main advantage of APD in relation to the PIN photodiode is a very high conversion coefficient. On the other hand, high inverse polarization voltages are necessary to reach the avalanche effect, a typical 200V for silicon APD (but 10-50 V for InGaSP), in relation to 5-20 V with the PIN diode. The noise generated in APD is increased, compared to PIN, and the response time is also increased (smaller bandwidth of APD) due to the passing time of the electrical charge carriers through the structure as well as the time necessary to provoke the avalanche effect. There are numerous improvements in the design of the avalanche photodiodes, which enable the bandwidth of over 5

GHz (together with the amplification-bandwidth product of 70 GHz).

There are a great number of types of optical receivers, but all have a photodiode as the basic element in common in which the conversion of the optical into the electrical signal is made. Two basic types of receivers are known: 1. for systems with intensity modulation and direct detection (IM/DD), and 2. for coherent systems. The receivers for coherent systems are more complex and varied. The receivers for IM/DD systems advanced very much and are still widely used with all new technological improvements which are the result of the development of coherent systems. It can be said, that IM/DD systems, by improving lasers and narrowing the range of the output signal, on the side of the transmitter, developed into coherent systems (in the sense of the source coherence) using different kinds of modulation, whereas on the receiver's side the concept of the direct detection remained practically the same (with additional improvements and changes, depending on modulation).

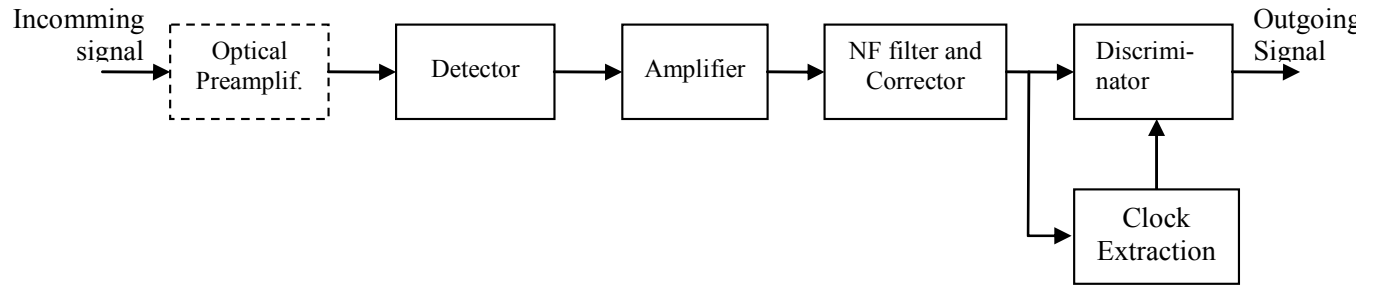

Figure 3. A typical block scheme of the optical receiver

The basic block scheme of a digital IM/DD receiver (hypothetically) is shown in "Fig. 3". The basic elements are detector (PIN diode or APD), amplifier, NF filter and equalizer (corrector), tact extractor and decision circuit. In many cases, the input signal is brought before the optical pre-amplifier before the detector, to improve the system performances, and with the newest optical components a semi-conducting optical pre-amplifier and a photodiode are realized in the same integrated circuit.

The main difference between the receivers in the coherent systems comparing to IM/DD ones is that the optical signal is brought out from the local oscillator to the photodiode, by the optical coupler, besides the input signal. The photodiode here has the role of the mixing circuit. If the frequencies of the input signal and the signal from the local oscillator are the same, it is a homodyne receiver, and at the output of the photodiode the signal appears translated into the basic range (only the upper side range). Another variant is a heterodyne receiver where the spectrum of the input signal is translated into the surrounding of the chosen mid-frequency. The mid-frequency signal is amplified at first, and then demodulated in the usual way, depending on the applied modulation type. The coherent receiver's sensitivity is bigger due to mixing of input and the signal from the local oscillator in comparison with IM/DD receiver for the multiplying factor $(P_l/P_s)^{1/2}$, where $P_l$ and $P_s$ are average powers of the local oscillator and the input signal. This fact is important as $P_l$ is significantly higher than $P_s$. On the other hand, the sensitivity is getting reduced due to the increased quantum noise which is now proportional to the increased photodiode current, owing to the larger signal from the local oscillator. That is why the correct comparison of sensitivity is only possible when the signals at the input of the coherent and "non-coherent" receivers are compared, when the bit error rate is the same.

## IV. OPTICAL AMPLIFIERS

There are a great number of optical amplifiers which can be classified into three groups: 1. with semiconductors, 2. those on the basis of the optical fibre, and 3. the amplifiers using non-linear effects.

Semiconductor Optical Amplifier (SOA) is similar to laser diode in its structure, and the amplifying is based on laser effect. There are two types of these amplifiers: with Fabry-Perot resonator (FPA), and with the traveling wave (TWA), the difference being that with FPA the amplifying is in numerous light passes through active zone, and with TWA the input light is amplified with only one passing. FPA has much narrower bandwidth in relation to TWA, but requires lower exciting current.

The most widely-spread type of amplifier used today is the amplifier based on the optical fibre doped by erbium (EDFA). EDFA are very good amplifiers, especially in DWDM systems which require a flat amplifying characteristic in wider range around 1550 nm. The work principle of this amplifier is shown in "Fig. 4a". A signal at wavelength of 820, 940 or 1480 nm (depending on choice) is pumped into the fibre with an input signal, over the directed coupler, from a relatively powerful laser diode. The signal excites erbium ions in the doped fibre which is 10-20m long. When there is no input signal, a spontaneous emission of non-coherent light appears, and thus newly formed photons further stimulate the emission of new photons along the doped fibre. In that way, the amplified spontaneous emission appears (ASE). When the input signal is present, the electrons excited by pump signal, from the energy level $E_3$ (where their average life time is about 1 μs) pass to the level $E_2$ (with average life time of 10 ms), from where they are returned to the basic level $E_1$ by the process of stimulated radiation, under the influence of the input signal, with the emission of the amplified signal. The amplifications which are achieved are over 30 dB.

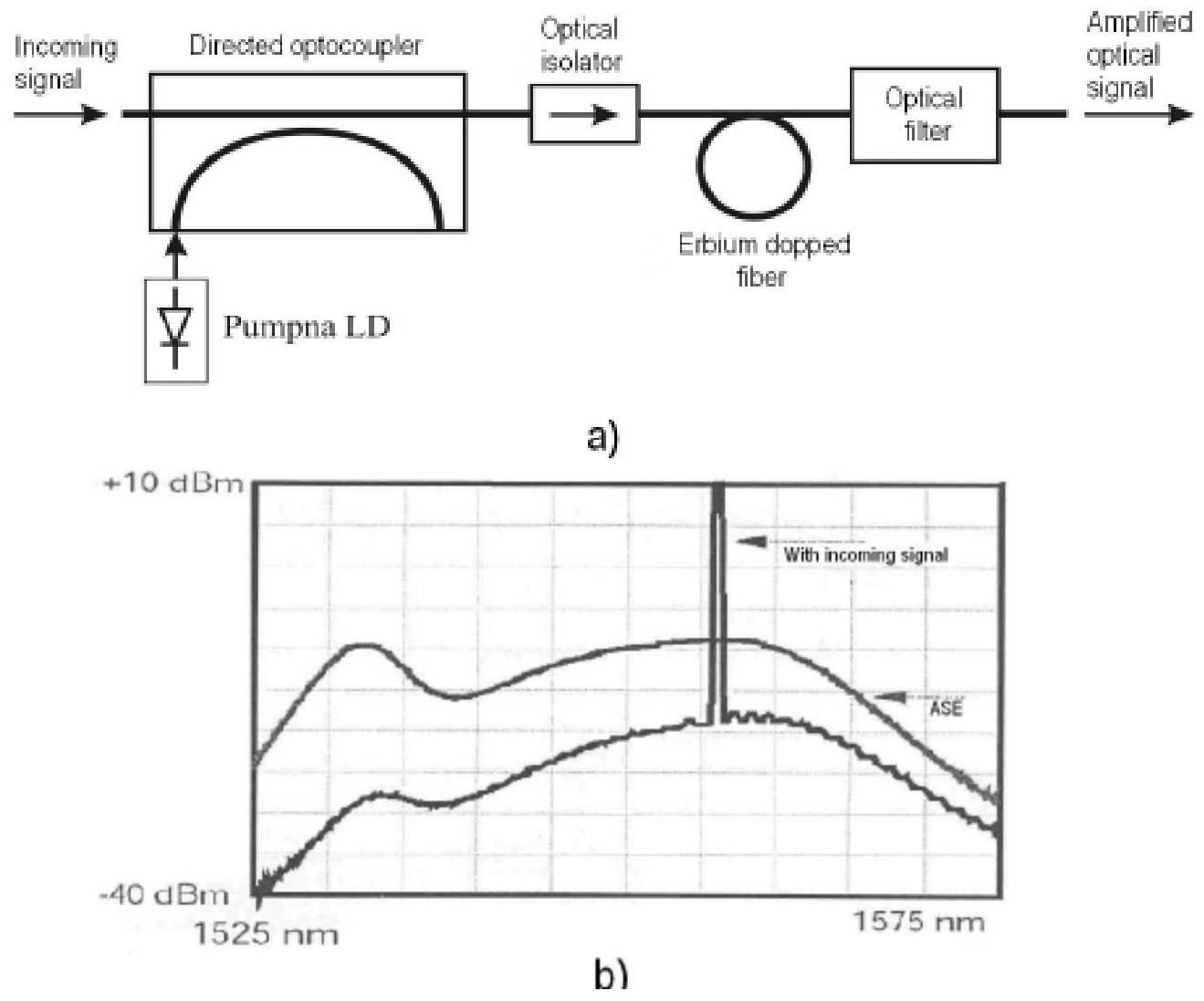

Figure 4. a) EDFA work principle; b) The response of EDFA amplifier in presence (lower curve) and in absence of the input signal (upper curve)

"Fig. 4b" presents the response of EDFA amplifiers in cases when there is and there is not the input signal that is to be amplified. Additional "flattening" filters are used in today's EDFA amplifiers, to gain the flat amplifying characteristic.

Apart from EDFA, there are also amplifiers for work at 1300 nm, with fibres doped by praseodymium. These amplifiers have lower gain and are less developed.

Among the amplifiers using non-linear effects in fibre the most famous is Raman's amplifier. The pump laser at shorter wavelength than the input signal is used here. Owing to the non-linear, so-called Raman's effect, the photons of the "pumped-in" signal scatter and lose energy whose one part convert into photon energy at the wavelength of the input signal, which causes the amplifying of the input signal. In that way, a signal at any wavelength can be amplified, but it is necessary for each wavelength of the input signal to have the adjusted wavelength of the pump laser, which is unfavorably with DWDM systems.

Optical amplifiers can be used at different places in the system, and can be:

- Output, i.e. power boosters which are placed at the output of the transmitter,
- In-line amplifiers which raise the level of the signal along the fibre,
- Pre-amplifiers, i.e. low-noise amplifiers before the receiver.

Typical characteristics of the basic types of amplifiers are shown in "Table 1".

TABLE I. A REVIEW OF THE CHARACTERISTICS OF THE AMPLIFIER'S BASIC TYPES

| | **LD** | | **Fibre** | **Non-linear effects** |
|---|---|---|---|---|
| | ***TWA*** | ***FPA*** | ***EDFA*** | ***Raman*** |
| Amplification | ~ 30 dB | 20-30 dB | 30-45 dB | 20-45 dB |
| Bandwidth | ~ 1000 GHz | 1-10 GHz | ~ 500 GHz | ~ 3000 GHz |
| Saturation power | ~ +10dBm | -10 ~ -5 dBm | +5 ~ +10 dBm | ~ +30 dBm |
| Polarisation dependance | yes | yes | no | no |
| Type of exciting | Current (100-200 mA) | Current (~ 20 mA) | LD source (20 - 100 mW) | Laser (1 - 5 W) |

## V. MODULATION OF THE OPTICAL SIGNAL

Two ways of modulations are possible in the optical transmitter: direct modulation of the laser current or light modulation after coming out of the laser into the outer modulator ("Fig. 5a"). The reason for using the outer modulators is a more effective control of the output signal characteristics. For systems at 40 Gb/s rate, the only possible way of modulation is by using outer modulators. In modern optical transmitters, two types of modulators are used: Mach-Zehnder interferometer (MZI), and electro-absorbtion modulator (EAM).

Figure 5. a) Two modulation ways: using the outer modulator; b) Mach-Zehnder interferometer

The MZI work principle is based on a strong electric-optical effect of the substrate made of lithium-neobate ($LiNbO_3$) or galium-arsenide (GaAs), where there is a change of refraction index of material due to the change of the electric field. One of the MZI realizations is shown in “Fig. 5b”, where there is a wave guiding structure with two Y-couplers. In this system, the electro-optical actuation cause the phase difference of $n \cdot \pi$ into two interferometer branches, so that at the end of the right Y- coupler there is an amplitude addition or annulment, and in that way, the phase modulation is turned into the amplitude one. MZI modulators are a key component in realizing different types of modulation.

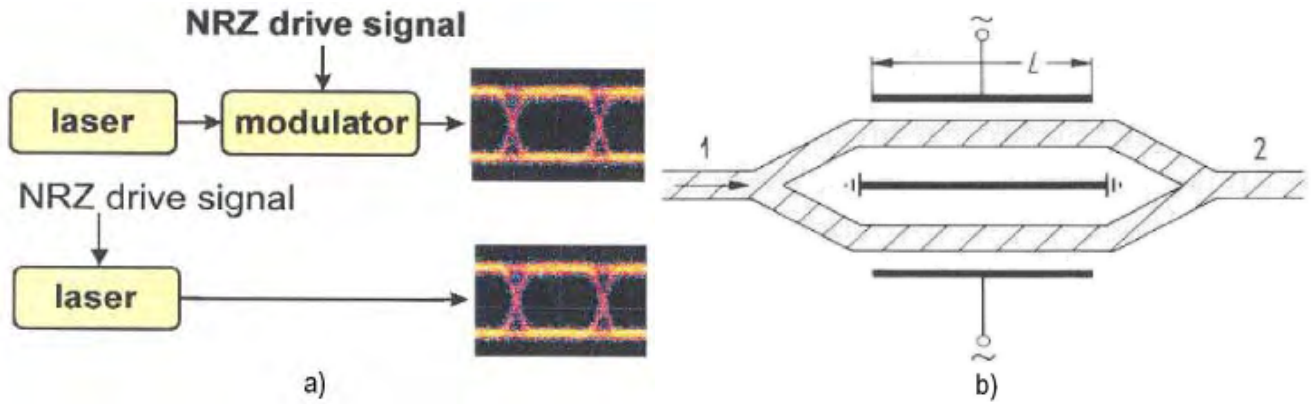

### A. *Data Modulation Types in Optical Systems*

Various data modulation formats are used optical systems so as to exploit the spectrum in the best possible ways and overcome the problems related to noise, dispersion and non-linear effects in the fibre.

The basic division is into the amplitude and phase modulation, the difference being that an additional phase modulation can be applied to the amplitude modulation (e.g. a chirped or a duobinary format) having no information. Likewise, there are phase modulations with additional amplitude modulation, e.g. return-to-zero phase modulation (RZ). With phase modulation formats, it is necessary to apply differential phase modulation, as direct detection (proportional to the optical power) is most often used at the receiver. As there is no referent signal as in "coherent" receivers, phase differences between neighboring bits are converted at the receiving side before the very detection into intensity modulated signal by means of interferometer with one bit delay (in one of its branches).

Most of modulation formats can be realized in two ways: with return to zero (RZ) and with no return to zero (NRZ). In realization of the RZ format, an additional modulator is usually used (pulse carver) of MZM or EAM type, as shown in “Fig. 6c” for the case of binary amplitude modulation (*Amplitude Shift Keying*-ASK or *On/Off Keying*-OOK).

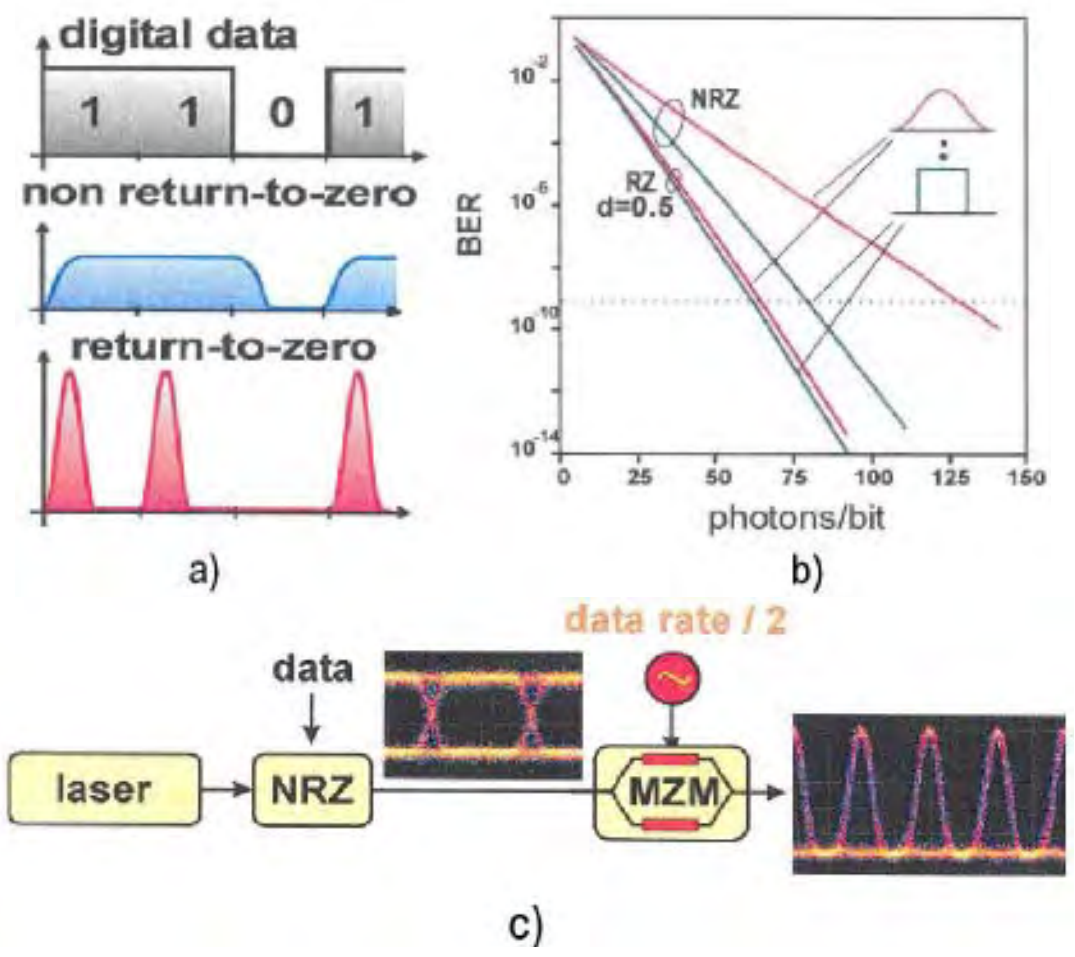

Figure 6. a) RZ and NRZ impulse width in relation to bit interval; b) BER depending on the energy and shape of impulse at the modulator output; c) RZ signal generating by the additional outer modulator

RZ formats are generally more tolerant to inter-symbol interference (ISI) which appears as the result of the limited optical and electrical bandwidth of the system. With RZ format, it is possible to achieve the increased sensitivity of the receiver of 2-3 dB in comparison to NRZ, with no necessary equalization. RZ format is also more tolerant to other effects in propagation through the fibre, such as: non-linear effects, polarization mode dispersion (PMD) or multi-path interference (MPI). Another advantage of RZ compared to NRZ format is much less dependence on the impulse shape, which is shown in “Fig. 6b”. On the other hand, the spectrum width is broadened, namely, the spectral RZ signal efficiency is less.

The applied data modulation format can also be "chirped" if a controlled amount of the analogue phase modulation is additionally applied. In the case of CRZ (chirped RZ) or alternate-chirp RZ (ACRZ) modulation, the signal spectrum width is broadened, which generally increases the resistance to non-linearity in the fibre, but with the reduced spectral efficiency. CNRZ format is used to increase the distance limited due to chromatic dispersion (e.g. from 60 to 140 km with systems of 10 Gb/s).

The most often used is binary OOK modulation where information bits are mapped in two optical levels. Among the binary modulations a binary differential phase modulation is also used (DPSK) whose sensitivity of the receiver is about 3 dB higher compared to OOK, due to higher tolerance to the noise level. DPSK allows bit rates of 40 Gb/s at very long distances.

In order to increase the digital rate when the symbol rate is the same, multilevel modulation formats are used. An example of this type of modulation is differential square phase modulation (dQPSK) where the bit pairs {00, 01, 10, 11} are mapped into the phase shifts {0, +π/2, -π/2, π} of the optical carrier. This modulation enables savings in needed bandwidth and increases tolerance to chromatic dispersion (CD) and PMD. On the other hand, the transmitter and the receiver are more complex.

One of the formats to increase spectral efficiency, distance and tolerance to non-linearity is amplitude modulation with suppressed optical carrier and return to zero (*Carrier-Suppressed* RZ-CSRZ). It can be said that it is, in fact, OOK modulation with alternate phase change by $\pi$ in each bit interval. The realization is different from OOK only in value range of the input impulse in outer MZM carver from Fig. 6. This format belongs to the group of so-called pseudo-multilevel modulations where the set of the output symbols is {0, +1, -1}. The advantages that are gained are the consequence of a more suitable shape of spectrum of signal.

Similar to CSRZ, so-called formats with partial response have a set of output symbols {0, +1, -1} where compression of the optical spectrum is achieved by the appropriate way of coding. The most important in this group of formats are duo-binary with no return to zero (NRZ-DB), and the format with alternate mark inversion and return to zero (RZ-AM).

Finally, single-sideband formats should be also mentioned (SSB) as well as vestigial sideband format (VSB). These formats are realized by optical filtering of OOK signal on the side of the transmitter or the receiver. These formats are used in DWDM because of its small width.

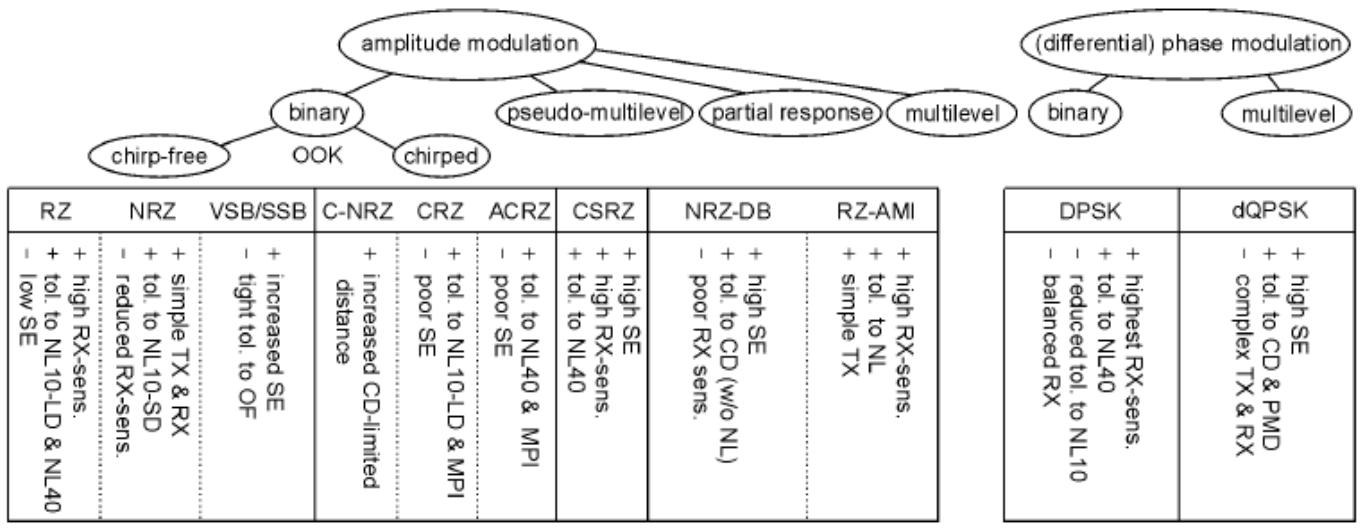

Figure 7. Review and characteristics of some most important modulation formats in optical transmission systems

"Fig. 7" presents the review of the most important optical modulation formats which are in use nowadays (with marked advantages and disadvantages), while the spectral shapes of some of them are shown in "Fig. 8".

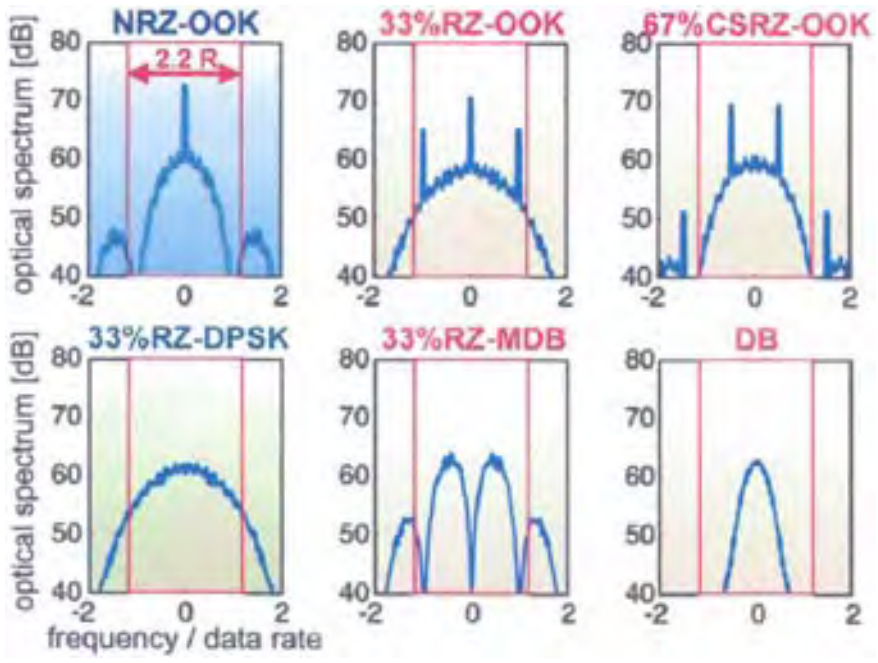

Figure 8. Spectrum shapes of some modulation formats and their width in relation to digital rate (R)

## VI. Current Status and Further Development of Optical Systems

In today's optical networks, data are transmitted in a standard way at rates of 2.5 Gb/s and 10 Gb/s and with the use of SDH technology. Many producers frequently test systems of 40 Gb/s rate, and some of them are commercially available. Some data from reference books [5] on maximum capacities and distances among today's optical systems are shown in Tables 2, 3 and 4. The first table refers to up to now installed systems, the second table comprises data on products which can be found in the market, while the third one relates to the results achieved in the experiments.

TABLE II. MAXIMAL ACHIEVED CAPACITIES AND DISTANCES IN SO FAR INSTALLED SYSTEMS

| Capacity [Tb/s] | Distance | Spectral efficiency | Modulation format | Capacity x distance |
|---|---|---|---|---|
| 0.2 (80x2.5Gb/s) | 800 km | 0.05 b/s/Hz | NRZ | 0.2 Pb/s km |
| 0.8 (80x10Gb/s) | 600 km | 0.2 b/s/Hz | NRZ | 0.5 Pb/s km |
| 0.09 (9x10Gb/s) | 2,900 km | 0.1 b/s/Hz | RZ (soliton) | 0.3 Pb/s km |
| 0.36 (36x10Gb/s) | 6,500 km | 0.3 b/s/Hz | CRZ | 2.3 Pb/s km |

TABLE III. CHARACTERISTICS OF COMMERCIALLY AVAILABLE SYSTEMS

| Capacity [Tb/s] | Distance | Spectral efficiency | Modulation format | Capacity x distance |
|---|---|---|---|---|
| 0.8 (80x10Gb/s) | 1000 km | 0.2 b/s/Hz | NRZ | 0.8 Pb/s km |
| 1.0 (96x10Gb/s) | 1600 km | 0.4 b/s/Hz | NRZ | 1.6 Pb/s km |
| 0.4 (40x10Gb/s) | 4800 km | 0.2 b/s/Hz | NRZ / RZ | 1.9 Pb/s km |
| 1.1 (112x10Gb/s) | 1200 km | 0.1 b/s/Hz | NRZ | 1.3 Pb/s km |
| 2.6 (64x40Gb/s) | 1000 km | 0.4 b/s/Hz | CSRZ | 2.6 Pb/s km |
| 1.0 (96x10Gb/s) | 9000 km | 0.3 b/s/Hz | CRZ | 8.6 Pb/s km |

TABLE IV. THE RESULTS ACHIEVED BY EXPERIMENT

| Capacity [Tb/s] | Distance | Spectral efficiency | Modulation format | Capacity x distance |
|---|---|---|---|---|
| 10.2 (239x42.7Gb/s) | 300 km | 0.64 b/s/Hz | VSB-NRZ | 3 Pb/s km |
| 6.4 (150x42.7Gb/s) | 2,100 km | 0.80 b/s/Hz | duo-binary | 13 Pb/s km |
| 6.4 (150x42.7Gb/s) | 3,200 km | 0.80 b/s/Hz | RZ-DPSK | 20 Pb/s km |
| 5.1 (119x42.7Gb/s) | 1,280 km | 0.80 b/s/Hz | CSRZ | 7 Pb/s km |
| 2.6 (211x12.3Gb/s) | 11,000 km | 0.26 b/s/Hz | CRZ | 28 Pb/s km |
| 1.6 (37x42.7Gb/s) | 10,000 km | 0.40 b/s/Hz | RZ-DPSK | 16 Pb/s km |
| 1.0 (23x42.7Gb/s) | 320 km | 1.00 b/s/Hz | VSB-CSRZ | 0.3 Pb/s km |
| 0.04 (1 x 40Gb/s) | 1,000,000 km | one channel | RZ | 40 Pb/s km |

Supposing the transmitting potential of the single-mode optical fibre is theoretically up to 50 Tb/s, it is obvious from the previous tables that there are still space for increasing capacity and further improvement of transmission systems. The increase of the transmitting capacity can be done by increase of digital rate or by channel multiplexing in Wavelength-Division Multiplex (WDM). The possibilities of optical transmission in spread spectrum (OCDM) are also considered. Maximal rates per channel realized by one laser are 40 Gb/s, namely 42.7 Gb/s, which practically reaches the maximal limit (on today's development level of electronic circuits). Recently, there have been more and more experiments with Optical Time-Domain Multiplex (OTDM), where there are rates even up to 320 Gb/s per channel. Multiplex process is being done here of a great number of optical signals with rates of 10 or 40 Gb/s, generated by a number of various sources of the same wavelength. The multiplex is done by optical components (delay lines, semi-permeable mirrors).

Nowadays, the most effective way for multiple capacity increase of the optical fibre is by using Dense-WDM technology (DWDM), where the multiplex of even up to 160 channels is achieved, with minimal distance between the channels of 25 GHz. It is for sure that, the less distance is between the neighboring wavelengths in spectrum, the more prominent non-linear (and others') effects are, and, besides, better quality lasers are needed. Even with 50GHz distance, the maximal rate per wavelength is about 10 Gb/s (for today's equipment). From technical and economical point of view, the obvious advantage of DWDM technology is a very high transmission capacity. As the demand for the bandwidth increases, it is possible to simply add new capacities by activating unused wavelengths.

At present-day level of development, DWDM technology is used on backbone point-to-point links in transport networks of big telecommunication companies. By using point-to-point DWDM links, the problem of the limited rate in network nodes is not solved, so data are still processed in electric domain. That is why, further development of DWDM in the direction of All-Optical Networks is to be expected, which will eliminate any electronically signal processing. This implies development of a series of follow-up technologies, such as: optical add-drop multiplexes (OADMs), optical switches (OXCs) and even optical routers, optical amplifiers of appropriate characteristics, etc.

There are two concepts of transition to all optical networks. One direction is network development with wavelength routing, where the communication between the any two nodes is done by optical channels (OCh) through the network from end to end. Another attitude is development of the optical packet network where commutation of optical packets is done, similar to the existing "electronic" packet networks (ATM, Frame Relay, and IP). In the last case, the optical case would comprise "the optical header" with lower digital rate (so that electronic circuits could have enough time to analyze it and route the packet), and from "optical payload" with higher rate, routed in OXC devices on the basis of the received header.

## VII. CONCLUSION

Modern optical telecommunication systems have to provide high bit-rates as 40 Gb/s and more at even longer distances. Each breakthrough in increase of rate and/or distance requires overcoming of a large number of impairments which appear. This means that all the components, including optical receivers, must be improved in order to satisfy more strict demands. New formats of optical signal have also been developed and the realization of its modulation and demodulation is very close related with the design of optical transmitters and receivers. Key elements of such designs are DFB lasers, PIN and avalanche photodiodes and Mach-Zehnder interferometers. Differential PSK modulation format (with return-to-zero) is main candidate in new applications and tends to replace traditional intensity modulation with On-Off-Shift-Keying, but also a number of other data modulation formats can find its place.